# Quantum geometry in low-energy linear and nonlinear optical responses of magnetic Rashba semiconductor (Ge,Mn)Te

Tsubasa Takagi[1,2,*], Hikaru Watanabe[3], Ryutaro Yoshimi[4], Yuki Sato[2], Shingo Toyoda[2], Atsushi Tsukazaki[5], Kei S. Takahashi[2], Masashi Kawasaki[1,2], Yoshinori Tokura[1,2,6], and Naoki Ogawa[2,7]

[1]Department of Applied Physics, University of Tokyo, Bunkyo-ku,
Tokyo 113-8656, Japan
[2]RIKEN Center for Emergent Matter Science (CEMS),
Wako, Saitama 351-0198, Japan
[3]Department of Physics, University of Tokyo,
Bunkyo-ku, Tokyo 113-8656, Japan
[4]Department of Advanced Materials Science, University of Tokyo, Kashiwa 277-8561, Japan
[5] Institute for Materials Research, Tohoku University, Sendai,
Miyagi 980-8577, Japan
[6] Tokyo College, University of Tokyo, Bunkyo-ku,
Tokyo 113-8656, Japan
[7] RIKEN Baton Zone Program (BZP), Wako, Saitama 351-0198, Japan

* Corresponding author. Email: tsubasa-takagi@g.ecc.u-tokyo.ac.jp

*Contact author: tsubasa-takagi@g.ecc.u-tokyo.ac.jp

**ABSTRACT**. Quantum geometry appears as a key factor in understanding the optical properties of quantum materials, with the anticipation on diverging or quantized responses near the Dirac and Weyl points. Here we investigate linear and nonlinear optical responses—optical conductivity and injection current— in a magnetic Rashba semiconductor in the mid-infrared region, with varying the Fermi energy across the Dirac point. We reveal that the linear optical conductivity reflects quantum metric, which remains finite irrespective of the diminishing joint density-of-states at lower photon energy. It is also confirmed that the magnetic injection current enhances depending on the energy of the Fermi level relative to the Dirac point. These optical spectra are nicely reproduced by our theoretical calculations with geometrical effects taken into account.

## I. INTRODUCTION.

Optical spectroscopy provides powerful means for detecting the topological and geometrical nature of solids along with electrical transport measurements [1–6]. Recent theories have pointed out that quantum geometry is encoded in both linear and nonlinear optical responses; for instance, the quantum metric, the real part of the quantum geometric tensor, is directly related to the optical conductivity through the dipole-transition amplitude [5–8].

The bulk photovoltaic effect (BPVE), which refers to the generation of charge current in the bulk of noncentrosymmetric materials without an electric bias, is one of the nonlinear optical processes and also characterized by the quantum geometry [5,9,10], as exemplified by Weyl node chirality [11] and Berry curvature dipoles [12]. Moreover, quantum metric and Christoffel symbols, which are relatively unexplored geometric quantities, are tightly related to the BPVE under certain symmetry constraints [5,10]. In addition to such fundamental interests, the BPVE attracts growing attention in the field of photo-sensing owing to its potential broadband and ultrafast responses [13]. It has been theoretically predicted and partially demonstrated that the BPVE is significantly enhanced when excitation occurs near the Dirac point, owing to the divergence of these geometric quantities [13,14].

Breaking of space-inversion symmetry brings about spin-splitting in the electronic bands under spin-orbit interactions. The Rashba-type spin-splitting is a typical example that leads to numerous exotic phenomena including the Edelstein effect, nonreciprocal transports [15,16], and possible topological superconductivity [17]. Dirac point at the crossing of these spin-split bands emerges with a divergent quantum metric, similar to the case of Berry curvature at the Weyl point [10,18,19]. In such spin-split bands, illumination by circularly polarized light drives the photocurrent due to asymmetric carrier distribution in $k$-space, so-called circular photogalvanic effect (CPGE) as one of the BPVE processes governed by the Berry curvature, imaginary part of the quantum geometric tensor [20,21]. The CPGE has been observed in bulk Rashba semiconductors [22,23], quantum wells [24], topological insulators [25], Weyl systems and so on [11,26]. Introduction of magnetism in these systems further breaks time-reversal symmetry. The Zeeman effect modifies the electronic states asymmetrically in $k$-space, which allows for the other types of BPVE; the magnetic injection current as formulated by using quantum metric [5,10,27]. When the system has a magnetization $\boldsymbol{M}$ normal to its polarization $\boldsymbol{P}$, photocurrent emerges by unpolarized light along the direction of toroidal moment $\boldsymbol{T} = \boldsymbol{P} \times \boldsymbol{M}$ from the viewpoint of symmetry, which can be controlled by the external magnetic field.

In this paper, we employ infrared reflection and photocurrent spectroscopies on thin films of magnetic bulk Rashba semiconductor (Ge,Mn)Te. The doped Mn atoms lead to a strong coupling between the spin-split bands and the external magnetic field, leading to the asymmetric band shift. Along with the tunability of the Fermi level by the sample growth parameters, the (Ge,Mn)Te films act as a versatile platform, distinct from TaAs or other Weyl systems, for investigating optical responses reflecting the quantum geometric quantities. We observed enhanced linear optical conductivity and unique photocurrent spectra in these films at mid-infrared photon energy down to 0.078 eV (~19 THz). Three samples with different hole densities exhibit distinct photocurrent spectra; two of them show increasing responsivity for decreasing incident photon energy. The theoretical joint-density-of-states (JDOS), which does not involve quantum geometric effects, shows a sharp reduction below 0.2 eV, in stark contrast to the enhanced optical conductivity and injection current observed experimentally, suggesting the critical role of such quantum contributions.

## II. RESULTS

The quantum geometrical effect is encoded in the linear optical conductivity $\sigma^{ab}(\omega) = \sigma_1^{ab}(\omega) + i\sigma_2^{ab}(\omega)$ through the following equation [6,8,28]

$$\sigma^{ab}(\omega) = -ie^2\hbar \int \frac{d^3\boldsymbol{k}}{(2\pi)^3} \sum_{m,n} \frac{v_{mn}^a v_{nm}^b}{\epsilon_{km}-\epsilon_{kn}-\hbar\omega+i\gamma} \frac{f_{mn}}{\epsilon_{km}-\epsilon_{kn}}. \quad (1)$$

where $\epsilon_{km} - \epsilon_{kn}$ denotes the energy difference between bands $m$ and $n$ at the same $k$ point, while

*Contact author: tsubasa-takagi@g.ecc.u-tokyo.ac.jp

$f_{mn} = f_{km} - f_{kn}$ represents the difference of Fermi-Dirac distribution. $\gamma = 1/\tau$ is the phenomenological width associated with optical transitions. The off-diagonal velocity matrix elements are expressed in terms of the position matrix [2,29,30].

$$v_{mn}^c = \frac{i}{\hbar}(\epsilon_{kn} - \epsilon_{km})r_{mn}^c . \quad (2)$$

Accordingly, the linear optical conductivity provides insights into the quantum metric through $v_{mn}^a v_{mn}^a \propto r_{nm}^a r_{mn}^a = g_{mn}^{aa}$.

The parent compound $\alpha$-GeTe hosts a giant Rashba splitting in its bulk owing to the polar lattice distortion along [111] and the large spin-orbit coupling [31]. Doping Mn atoms to the Ge site keeps the lattice distortion, while their spins align ferromagnetically along [111] via the Ruderman-Kittel-Kasuya-Yosida (RKKY) interaction [32,33]. Therefore, the ground state of (Ge,Mn)Te becomes a multiferroic semiconductor as revealed by transport and soft-X-ray ARPES measurements [33–35]. We use the experimental coordinates $x$, $y$, and $z$ as shown in Fig. 1(a) in the following. The Hamiltonian representing the Rashba system with magnetization $M$ is

$$H = \frac{\hbar^2 k^2}{2m^*} + \alpha_R(k_x \sigma_y - k_y \sigma_x) + \Delta_z(\hat{M} \cdot \sigma) , \quad (3)$$

where $m^*$, $\alpha_R$, $\sigma = (\sigma_x, \sigma_y, \sigma_z)$, $\Delta_z$ and $\hat{M}$ are the effective mass, the Rashba coupling, the Pauli matrices, the Zeeman gap and the unit vector pointing to magnetization [16,33,36]. Here, we assume $\Delta_z = 50$ meV [33,34], which opens the mass gap at the Dirac point with $\hat{M} \parallel z$ (Fig. 1(b)). In the following, we discuss the electronic structure of the valence band under hole doping. We apply an in-plane magnetic field to orient the magnetization $\hat{M}$ along the $x$-axis. Accordingly, the mass gap closes, and the Dirac point shifts by $-\Delta_z/\alpha_R$ from $Z$ point (Fig. 1(c)). Therefore, optical excitation becomes asymmetric along the toroidal moment $T \parallel y$, which activates the magnetic injection current. The magnetic injection current conductivity is given by [37,38]

$$\sigma_{inj}^{abc}(\omega) = \tau_D \frac{\pi e^3}{2\hbar^2} \times$$
$$\int \frac{d^3 k}{(2\pi)^3} \sum_{m,n} f_{mn} \Delta_{mn}^a \{r_{nm}^b, r_{mn}^c\} \delta(\omega_{mn} - \omega), \quad (4)$$

where $\tau_D$ is the relaxation time and $\hbar\Delta_{mn}^a = v_{mm}^a - v_{nn}^a$ the interband transition of the group velocity. The band-resolved quantum metric $\{r_{nm}^b, r_{mn}^c\} = r_{nm}^b r_{mn}^c + r_{nm}^c r_{mn}^b = 2g_{mn}^{bc}$ represents the dipole-transition amplitude. The JDOS $\int [dk] f_{mn} \delta(\omega_{mn} - \omega)$ also governs the overall amplitude of the injection current conductivity.

The (Ge,Mn)Te films with a thickness of approximately 75 nm were grown on InP(111)A substrates after depositing a few monolayers of $Sb_2Te_3$ and GeTe buffer layers by molecular beam epitaxy (MBE) [16]. The hole density $p$ was controlled by the supplied amount of Te. The films were capped with $AlO_x$ by atomic layer deposition to prevent degradation. The Mn composition was determined to be 9% by inductively coupled plasma mass spectroscopy measurement.

We estimated the hole density $p$ of the (Ge,Mn)Te films by the normal Hall effect (Fig. 1(d)), and Curie temperature $T_c$ by the anomalous Hall effect [16] as summarized in Table I. Three samples, two with the Fermi energy below and one above the Dirac point, are examined in the following. Figure 1(e) shows the temperature dependence of the longitudinal resistance. It is seen that the longitudinal resistance increases with decreasing the carrier density $p$. The sample with hole density $p = 1.8 \times 10^{19}$ cm$^{-3}$ shows a broad upturn at low temperatures, consistent with localization effects as discussed previously [32,39].

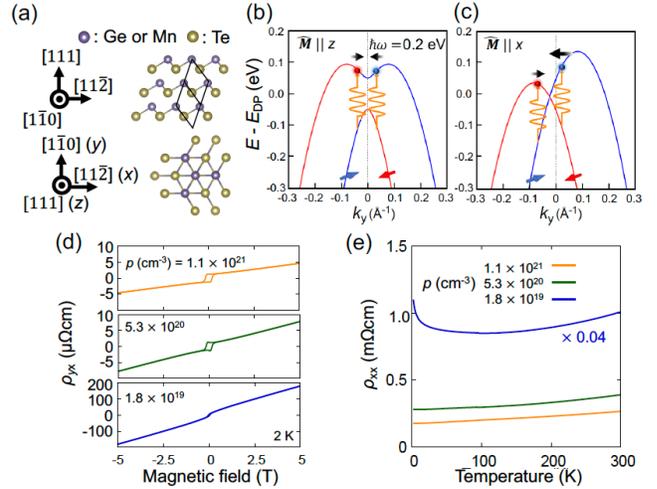

FIG. 1. (a) Schematics of rhombohedral crystal structure of (Ge,Mn)Te. In the main text, [111] and [11$\bar{2}$] are denoted as $z$- and $x$-axis, respectively. (b)(c) Schematic spin-split band structure in the ferromagnetic state, where magnetization points to $z$ direction (b) and $x$ direction (c). In-plane magnetic field induces an asymmetry in the electronic bands without the mass gap of Dirac electrons found in (b). (d) Magnetic field dependence of Hall resistivity $\rho_{yx}$ at 2 K for three samples with different hole concentrations. (e) Temperature dependence of longitudinal resistivity $\rho_{xx}$ measured without an external magnetic field.

*Contact author: tsubasa-takagi@g.ecc.u-tokyo.ac.jp

TABLE I. Sample characteristics. Considering the experimental hole density $p$, the $E_F$ values are evaluated based on *ab*-initio calculation in the presence of Zeeman field of $h_x = 50$ meV and with $E_{DP} = 6.381$ eV. $T_c$ is deduced from the temperature dependence of $\rho_{yx}$.

| hole density $p$ (cm$^{-3}$) | $1.1 \times 10^{21}$ | $5.3 \times 10^{20}$ | $1.8 \times 10^{19}$ |
|---|---|---|---|
| $E_F - E_{DP}$ (eV) | -0.180 | -0.073 | 0.180 |
| $T_c$ (K) | 150 | 120 | 20 |

Our experimental setup is schematically shown in Fig. 2(a). While the left part of the InP substrate is covered by the (Ge,Mn)Te film, the right part, shown in grey in Fig. 2(a), was clamped during preparation and has no film deposited. We performed real-space photocurrent mapping for the area indicated by a red rectangle in Fig. 2(a) and confirmed that the spontaneous photocurrent appears uniformly. Thus photo-thermal and Schottky effects around the electrodes were found to be negligible. It should be noted that the photocurrent is influenced by the impedance of cables and the preamplifier in the circuit. We discuss the photocurrent signal at the center (spot position (0,0) in Fig. 2(a)) hereafter.

Figure 2(b) displays the time-trace of the photocurrent averaged over 150 pulses under varying magnetic fields. The current peak at time zero is the signal we discuss. The subsequent structures around 5 ns are due to the impedance mismatch in the circuit. The photocurrent along the $y$ direction appears with the application of the magnetic field in the $x$ direction, consistent with the generation of magnetic injection current for the Rashba-type spin splitting in GeTe [40,41]. Note that the symmetry breaking at the surface and trigonal warping of electronic bands in the bulk are not essential for the magnetic injection current [42].

Figure 2(c) shows the magnetic field dependence of the injection current, excited at photon energy of 0.117 eV (10.6 μm in wavelength). The injection current increases with the external magnetic field up to +3 T and then saturates, which nearly follows the in-plane magnetization of the sample. Figure 2(d) depicts the injection current responsivity under the magnetic field of +5 T as a function of the temperature, whose amplitude rapidly reduces with increasing temperature and vanishes around 40 K.

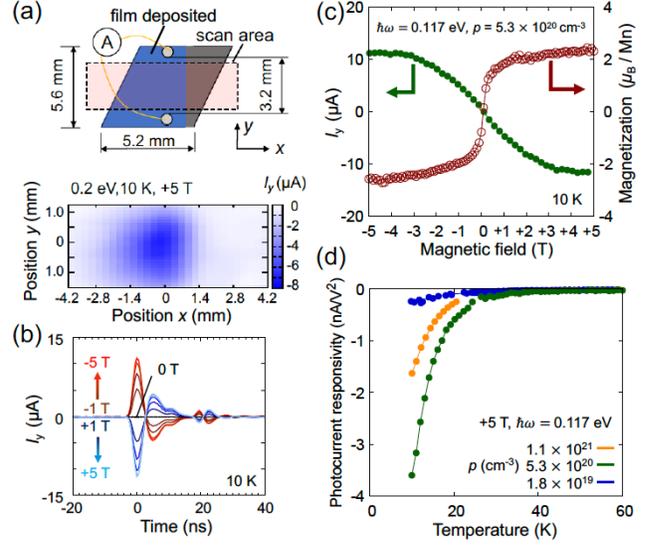

FIG. 2. (a) Schematics of photocurrent mapping with experimental coordinates. The zero-bias photocurrent in $y$ direction is measured under the in-plane magnetic field ($B_x$) at normal incidence of mid-infrared light. The blue shaded area indicates the part where the sample films are deposited. The laser spot was raster-scanned on the red-shaded area (top) for the mapping measurements, and the resultant photocurrent distribution is shown in the bottom. (b) The time traces of the photocurrent at 10 K. (c) Magnetic field dependence of photocurrent ($I_y$) for the sample with $p = 5.3 \times 10^{20}$ cm$^{-3}$, measured at 0.117 eV photon energy, plotted together with the magnetization under $B_x$. (d) Temperature dependence of photocurrent responsivity for three samples with varying hole concentrations.

Figure 3(a) presents the band diagram for GeTe obtained by *ab*-initio calculation. The Zeeman term $H_{zeeman} = \sum_i \mathbf{h}_{eff} \cdot \mathbf{\sigma}$ is added to incorporate the magnetic interaction ($|\mathbf{h}_{eff}| = 50$ meV). The $k$-space asymmetry of the bands with the Dirac point shifted in the $y$ direction is reproduced. Considering these band structures and density of states, we estimate the Fermi energy $E_F$ by using the experimentally determined hole density $p$ (Table I). It is seen that the sample with $p = 5.3 \times 10^{20}$ cm$^{-3}$ (green line) has the Fermi energy closest to the Dirac point at $Z$ point in the Brillouin zone. Below 0.2 eV, in which optical transition within the hole-doped valence bands dominates, the calculated JDOS between spin split bands (Fig. 3(b)) shows a monotonic decrease towards the lower photon energy. For the sample with the hole density $p = 1.8 \times 10^{19}$ cm$^{-3}$ (blue line), the finite JDOS originates from the contributions from away from $Z$ point.

*Contact author: tsubasa-takagi@g.ecc.u-tokyo.ac.jp

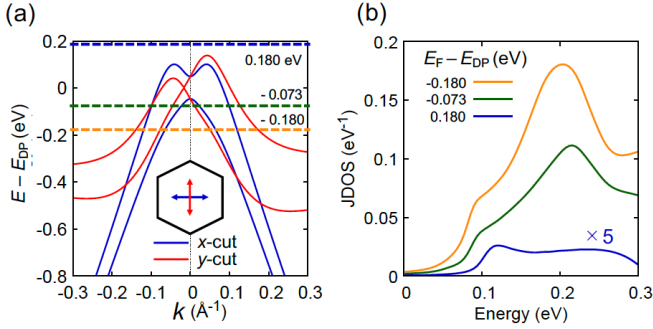

FIG. 3. (a) Band structure of GeTe obtained by *ab-initio* calculations in two representative directions. The Zeeman field of $h_x = -50$ meV is added. The Fermi energy $E_F$ is represented by dashed lines for each sample (in corresponding colors, blue: $p = 1.8 \times 10^{19}$ cm$^{-3}$, green: $p = 5.3 \times 10^{20}$ cm$^{-3}$, orange: $p = 1.1 \times 10^{21}$ cm$^{-3}$). The hexagon in the inset represents the $k_x$-$k_y$ plane containing Z point in the Brillouin zone. (b) Calculated JDOS under the Zeeman field of $h_x = -50$ meV for three samples with varying Fermi energy.

Linear optical conductivity offers valuable insights. We measured the optical conductivity of the (Ge,Mn)Te films from the normal-incidence reflectance by employing the Kramers-Kronig relations [43] Three spectra are plotted together with the Drude-Lorentz fittings in Figs. 4 (a)-(c) (see Table II for the fitting parameters). In contrast to the L1 and L2 peaks centered around 0.1 eV, the L3 exhibits qualitatively different features; monotonic increase toward higher energy and the short relaxation time $\tau_{L3}$. Figure 4(d) illustrates the sum of the L1 and L2 terms, which are compared to those from the *ab-initio* calculation (Fig. 4(e)). Here the contribution from intraband transitions (Drude components) are not taken into account. It is found that the calculated optical conductivity spectra (Fig. 4(e)) qualitatively reproduce the experimental ones (Fig. 4(d)).

From these analyses, we deduced experimental parameters for the evaluation of injection conductivity tensor; relaxation for $\tau_D$ in (Eq. 2) [44,45] and the Lorentzian width associated with optical transitions determined by using L1 and L2 with Matthiessen's rule ($1/\tau = 1/\tau_{L1} + 1/\tau_{L2}$). The linear conductivity, for both experimental and theoretical ones, reaches a maximum for the sample with $p = 5.3 \times 10^{20}$ cm$^{-3}$ (green line), which originates from the dipole-transition enhanced by the quantum metric integrated over the Brillouin zone.

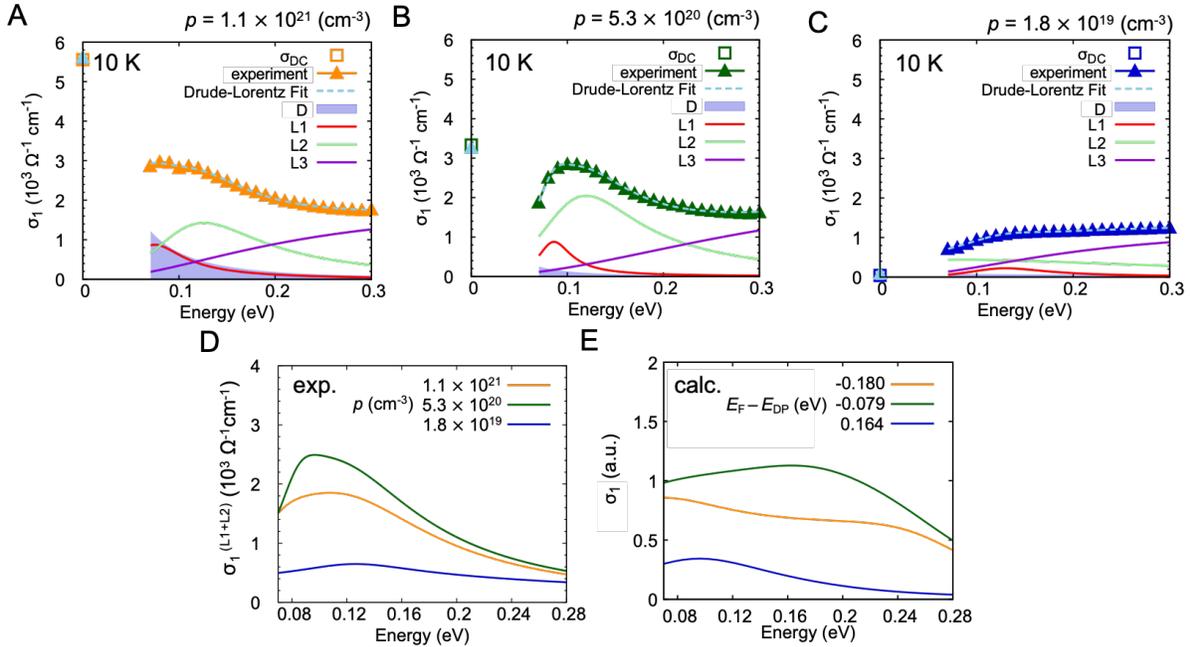

FIG. 4. (a)-(c) Optical conductivity spectra for three samples with varying hole concentration. The curve filled in blue represents Drude contributions (D: Drude term), while the colored solid curves are inter spin-split band components (L1, L2 and L3: Lorentz term) (d) The sum of the L1 and L2 terms extracted by the Drude-Lorentz fit. (e) Calculated real part of the optical conductivity

*Contact author: tsubasa-takagi@g.ecc.u-tokyo.ac.jp

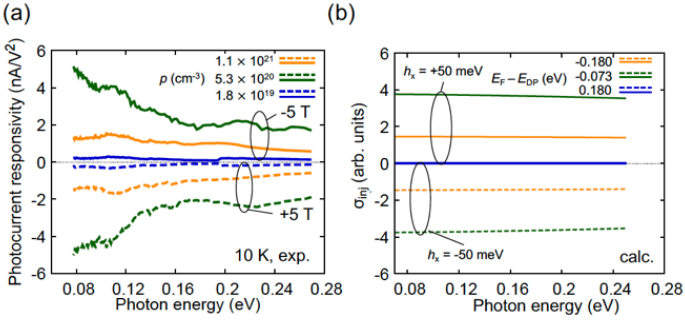

FIG. 5. (a) Photocurrent responsivity spectra for three samples with varying hole concentration. Dashed and solid lines represent the spectra under +5 T and -5T, respectively. Note that the photocurrent responsivity is underestimated by several orders in magnitude due to the impedance and parasitic capacitance in the circuit. (b) Calculated injection current conductivity spectra under the Zeeman field of $h_x = \pm 50$ meV.

Figure 5(a) presents the experimental injection current spectra in three samples down to 0.078 eV (∼19 THz). Two of them ($p = 5.3 \times 10^{20}$ and $1.1 \times 10^{21}$ cm$^{-3}$) show increasing responsivity as the photon energy decreases, while the one with the Fermi level above the Dirac point shows a rather small value. By using $\tau_D$ and Lorentzian width discussed above, we plotted the calculated injection current spectra in Fig. 5(b), which are found to align well with the experimental ones. The magnetic injection current at the lower energy is finite for both experimental and calculated spectra.

## III. DISCUSSION

The injection current conductivity is composed of three contributions (Eq. 2); interband transition of group velocity, JDOS, and quantum metric. The experimentally observed enhanced injection current at the lower photon energy signifies that the quantum metric plays a pivotal role, since the velocity does not diverge near the Dirac point, and the JDOS by the *ab-initio* calculation diminishes. Furthermore, we found that the observed enhancement becomes prominent when the Fermi energy comes close to the Dirac point [14], which is consistent with a recent model study for other Dirac materials [10]. Since there is no peak in both the optical conductivity and the injection current spectra near 0.2 eV, the density-of-state associated with the van Hove singularity has little effect (see Fig. 3(b)). It should be noted that the increase in the optical conductivity originates from the enhanced dipole-transition amplitude as described by the band-resolved quantum metric, particularly in the low-frequency region.

*Contact author: tsubasa-takagi@g.ecc.u-tokyo.ac.jp

Now we turn to the temperature dependence of the injection current (Fig. 2(d)). The injection current is generally temperature-dependent, owing to the changes in exchange coupling strength, relaxation time, and transition probability through the Burstein effect and so on [25,45,46] The dramatic increase in injection current at low temperatures can be attributed to the reduced phonon scatterings of the photoexcited carriers [47,48]. Note that the intrinsic Fermi surface effect, which is another photocurrent mechanism that shows a low-energy enhancement in itinerant systems, is independent of relaxation time and is expected to play a minor role in the present case [5,49,50].

## IV. CONCLUSIONS

To summarize, we studied linear and nonlinear optical responses, optical conductivity and magnetic injection current, in magnetic Rashba semiconductor (Ge,Mn)Te thin films in the mid-infrared region, with varying the Fermi energy across the Dirac point. At lower photon energy, the linear optical conductivity remains finite despite the diminishing JDOS, and the magnetic injection current also shows enhancement, depending on the relative energy of the Fermi level and the Dirac point. It is discussed that the quantum geometrical effects are necessary to reproduce these experimental optical spectra by theoretical calculations, unveiling their essential contributions in real samples.

## ACKNOWLEDGMENTS
This work was supported by RIKEN Junior Research Associate Program and the RIKEN TRIP initiative (Multi-Electron Group).

## APPENDIX

**Drude-Lorentz fitting parameters**
TABLE II. $\omega_p$ and $\tau_D$ are the plasma frequency and relaxation time for the Drude term. $\tau_{L1}, \tau_{L2}$ and $\tau_{L3}$ are the relaxation time for the Lorentz terms.

| hole density $p$ (cm$^{-3}$) | $1.1 \times 10^{21}$ | $5.3 \times 10^{20}$ | $1.8 \times 10^{19}$ |
|---|---|---|---|
| $\omega_p$ (eV) | 0.57 | 0.32 | 0.21 |
| $\tau_D$ (fs) | 17.7 | 32.2 | 1.89 |
| $\tau_{L1}$ (fs) | 9.64 | 15.5 | 6.32 |
| $\tau_{L2}$ (fs) | 4.60 | 5.02 | 1.97 |
| $\tau_{L3}$ (fs) | 0.18 | 0.18 | 0.31 |

**Linear optical and injection current spectroscopy**

We performed reflectivity measurements in the range of 0.07-0.99 eV by using JASCO FTIR6300 and a cold-finger type cryostat.

For the injection current spectroscopy, the AuPd electrodes were deposited along the *y* direction after removing the capping layer. The short-circuit photocurrent was detected using a wide-band preamplifier (100 MHz bandwidth) under the illuminated by a pulsed laser source (160 fs, 6 kHz, spot size ~1.1 mmφ) at normal incidence. The sample was placed inside a magnetic cryostat with Voigt geometry. We recorded time trace of the photoresponse averaged over 150 pulses under various magnetic fields.

### *ab*-initio calculation

The electronic structure of GeTe is obtained using the Quantum Espresso code [51,52]. Based on the density functional theory (DFT). We use the norm-conserving Vanderbilt pseudopotential [53,54] with the generalized gradient approximation of full-relativistic Perdew-Burke-Ernzerhof for the exchange-correlation functional [55]. The self-consistent band structure calculations are performed with the k mesh 7×7 ×7, and the plane-wave cutoff is 100 Ry for the wave functions.

The physical properties are evaluated with the Wannier functions [56], comprised of Ge: p-orbitals and Te: p-orbitals. After the projection onto the Wannier orbits, the Zeeman magnetic field of 50 meV is included to reproduce the exchange splitting from the dilute ferromagnetism of Mn atoms. JDOS, interband contribution to linear optical conductivity, (Eq. 2) and injection current response (Eq. 1) are evaluated with the discretized Brillouin zone meshed as $360^3$, $360^3$, and $288^3$, respectively. For numerical convergence, the scattering factor is introduced as $\delta(x) \to \pi \gamma(x^2 + \gamma^2)$. The scattering factor $\gamma$ is 10 meV for the calculations of JDOS and linear optical conductivity, and $\gamma = \left(\frac{1}{\tau_{L1}} + \frac{1}{\tau_{L2}}\right)^{-1}$ for that of injection current. The latter refers to the experiment values since the injection current response is sensitive to the scattering factor.

*Contact author: tsubasa-takagi@g.ecc.u-tokyo.ac.jp

*Contact author: tsubasa-takagi@g.ecc.u-tokyo.ac.jp

*Contact author: tsubasa-takagi@g.ecc.u-tokyo.ac.jp